\documentstyle[aps,preprint]{revtex}
\begin{document}
\title{Heavy-Fermions in a Transition-Metal Compound: $LiV_2O_4$}
\author{ C.M. Varma}
\address{Bell Laboratories, Lucent Technologies, Murray Hill, NJ 07974}

\maketitle

\begin{abstract}
 The recent discovery of heavy-Fermion 
properties in Lithium Vanadate and the enormous difference in its properties from
 the properties of
Lithium Titanate as well as of the manganite compounds raise some puzzling
questions about strongly correlated Fermions. These are disscussed as well
as a solution to the puzzles provided.
\end{abstract}

\newpage

\section*{Introduction}
 The properties of $LiV_2O_4$ (LiV)for $T\lesssim 20K$ are those of a heavy-Fermi-liquid \cite{kondo} \cite{takagi}: the specific heat $C_v \sim  \gamma T$ with $\gamma \approx
0.5 J/mole K^2$, Pauli susceptibility $\chi$  with $\chi T/C_v \approx 1.8$,   the resistivity $R(T) = R(0) + A T^2$ with $A \sim \gamma^2$ lying on the 
Kadawoski-Woods \cite{kada} plot. These parameters are similar to those in $UPt_3$
and many rare-earth compounds of Ce and Yb\cite{heavy}. This discovery raises some very interesting issues in our understanding of strongly correlated Fermions.

 $LiV_2O_4$ (just as $LiTi_2O_4$\cite{liT} (LiT)) has the spinel structure with 2 Transition metal
(TM) ions per unit-cell in {\em equivalent} sites. So, it is a mixed valent 
compound with equal ratios of $V^{3+}$, which has $S=1$ and $V^{4+}$ which
 has 
$S=1/2$. At first appearance the Hamiltonian of the system is similar to
the Jonkers-Van Santen compounds\cite{jV}, like $La_{2-x}Sr_xMnO_3$ (LMN),
 which are also mixed valent with ratio $x/(1-x)$ of $Mn^{3+}$ to
 $Mn^{4+}$. LMN for $x\sim 0.3$ is
a ferromagnetic metal, whose properties are well described by the Double-Exchange model. The first question is why does LiV behave 
so completely differently than (LMN)? Indeed when is the
 Double-Exchange model valid?

The isostructural neighbor to LiV, LiT is also mixed valent with
 equal ratio of $Ti^{3+}(S=1/2)$ and $Ti^{4+}(S=0)$. This is an 
ordinary metal with mass enhancement of $O(1)$. Why then the 
dramatic difference between TiV and LiV?

The bare hybridization parameters of rare-earth and actinide
 compounds are typically more than an order of magnitude smaller
 than the transition metal compounds. The effective mass observed
 for them is of the right order of     magnitude as arising from the Kondo-effect of the moments in f-orbitals. Assuming 
the mass renormalization in LiV is also a Kondo-effect, 
why is the effective mass similar to that in the rare-earth
 and actinide compounds?

A final question of-course is the applicability of the Kondo-effect 
and 
associated ideas to compounds like LiV with just one species of
 electrons. Such ideas have usually been applied to a lattice of 
(atleast) two
kinds of ions, one of which has f-orbitals with well-localised
 magnetic moments
(because the local correlation energy is much larger than the
 hybridization
energy with the neighbors) interacting with weakly interacting itinerant electrons.
 In LiV, the same-electrons act as local moments that are 
Kondo-quenched as well
as the electrons that do the quenching.

It is easiest to start with the final question. A mean-field method for  correlated Fermions on a lattice has been recently developed by
 considering the 
problem in the limit of large-dimensions\cite{vollhardt}, \cite{george}.        One of the most fruitful
 applications
of the method is to consider 1 ion in a bath whose properties 
(static as well
as dynamic) are determined self-consistently. For the one band
 Hubbard model,
for example, the Hamiltonian coupling the ion to the lattice is
 simply the Anderson model for local magnetic moments in which the 
parameters are determined self-consistently\cite{george}. From this point of
 view there is no
formal difference in treating the one-band Hubbard model or the 
multi-band models, with which Heavy-Fermions are customarily treated.
 While much remains yet to be developed,
especially in the question of effective-interaction between ions, the experimental results in LiV may be taken as further validation of
 this approach.
If we adopt this approach, the other questions in the Introductions 
may be addressed by considering the competition between the Kondo-effect quenching 
magnetic moments of an impurity embedded in itinerant electrons and the magnetic-interaction between ions favoring the mangetic moments.
 The difference between the pair of impurity problem and the actual 
lattice is then usually a
difference of numbers (which in practice is always less than an 
order of magnitude).

\section*{Lithium Vanadate and Lithium Titanate}

The difference of the properties of LiV (mixed valent with $S=1$) and
 $S=1/2$ and LiT (mixed valent with $S=1/2$ and $S=0$) is reminescent
 of the difference in properties of mixed valent rare-earth compounds
 of Ce and Yb on the one hand and of Tm on the other\cite{cmv} \cite{batlogg}. One of the valences of Ce($f^0$) and of Yb($f^{14})$)
is non-magnetic, while both valence states of Tm in TmSe etc. are
 magnetic 
(ignoring a small crystal-field splitting). The dominant interaction
 of mixed valent systems with Hund's rule energy comparable or larger
 than the hybridization energy is Double-exchange. If ${\bf S_i}$ is the moment
 of one of the valences and ${\bf (S + 1/2)_{j,max}}$ that of the other, then the Double-Exchange Coupling\cite {anderson}
between two ions i and j is
\begin{equation}
(t_{ij}/(2S+1))|{\bf S_i+S_j+1/2}|.
\end{equation}
If either ${\bf S_i}$ or $({\bf S+1/2)_{j,max}}$ is 0, there is no magnetic 
interaction to leading order. Moreover the effective Kondo-temperature 
for the mixed-valence problem is just the hybridization width. For the single band problem as in
LiT this is less than an order of magnitude smaller than the one-particle
bandwidth. This is much larger than any second-order magnetic interactions. This explains why LiT behaves as an ordinary metal with an effective mass enhancement of order unity; i.e. a specific heat coefficient $\gamma$ which is only a few 
$milliJoules/molecm^2$. The mixed-valent compounds of Ce and Yb have a
$\gamma$ of 50 to 100 $milliJoules/molecm^2$ because the bare hybridization parameters of f-electrons are smaller than those of d-electrons by a corresponding amount.

\section*{heavy-Fermion behavior of Lithium Vanadate}

Why then does LiV not exhibit the properties of the Double-exchange model 
and 
be ferromagnetic as LMN  and TmSe (when suffficiently mixed valent) do? The answer can be found in the energetics of the successive crossovers that
a $S\ne 1/2$ moment must undergo in the Kondo Effect. These can be estimated 
on the basis of variational calculations reported sometime ago\cite{yafet}.
The variational approach in such problems foreshadowed the so-called 
no-crossing approximation\cite{nocross}, the 1/N approximation\cite{1/N} and the
slave-Boson approximations\cite{slave}. The conclusions drawn here could be derived equally well by these methods.

The states of an (orbitally degenerate) mixed valent (S=1,S=1/2) impurity
in a metal can be a spin-triplet, a spin-doublet or a spin-singlet. the wavefunction for each of these states and their energy is given in Ref. (12).  In this case as well as the simpler $S=1/2$ problem, the Kondo-Temperature which sets the scale for the low temperature properties is the difference in the binding energy of the singlet and the doublet states. But for the mixed-valent V ion, one must also consider the energy difference of the triplet and the doublet state as well. This difference sets the scale for the crossover to an effective $S=1/2$ problem. The binding energy of the triplet state is
very small compared to that of the doublet and the singlet state, which are very close in energy. So the triplet state can be ignored. The binding energy of the 
doublet ($k_BT_D$) is of the order of the hybridization energy. The binding energy of the singlet ($k_BT_S$) is lower than that only by O($10^{-2}k_BT_D$).
The difference in binding energy for these states arise from the different phase-space for scattering allowed in each of the spin-states and has been fully explained in Ref. (12). 

Given these energies, it follows that for $T\lesssim T_D$, the properties of a single mixed valent Vanadium impurity are those of the S=1/2 problem until a $T = T_F$ of $O(T_D - T_S)$. Below this temperature the properties are that of a Fermi-liquid with an effective Fermi-
temperature $T_F$. If $T_D$ is much larger than the double-Exchange parameter,
Double-exchange is irrelevant and the (thermodynamic) behavior of the periodic lattice can be calculated from that of a single-site problem.

A reasonable number for the hybridization energy is $O(10^3)K$, i.e. an order of magnitude smaller than the one-electron bandwidth. Then below this temperature the property of the system is that of a $S=1/2$ problem. These calculations then explain why the heavy-fermion behavior occurs with $T_F$ of
about 20K as well as show that the effective magnetic moment above $T_F$ up
to a very high temperature is of S=1/2 rather than the mean of $S=1$ and $S=1/2$. 
Indeed, the magnetic susceptibility above $T_F$ and below 300K has the Curie-constant corresponding to $S=1/2$\cite{kondo,takagi}. This is a strong test of the ideas and
results presented here.

\section*{Double-Exchange in Lanthanum Manganite}

Finally we come to the question of why the Kondo-effect does not eliminate the 
possibility of ferromagnetism through Double-exchange in LMN. The reason has to do with the details of the electronic structure of the $Mn^{3+}$ and $Mn^{4+}$ ions. The latter has 3 Hund's rule coupled electrons in the $t_{2g}$ orbital
while the former has another electron Hund's rule coupled but in the $e_g$ orbitals. The ionization energy to go from the former to the latter is on the scale of $1eV$. While two ions are then degenerate when considering the energetics of 
Double-exchange, they are not mixed-valent for purposes of the energetics of the Kondo effect. The first stage in the Kondo-effect would be a crossover from a $S=2$
to a $S=3/2$. The effective exchange parameter for this is the square of the hybridization energy divided by the ionization energy, which is then an order of magnitude smaller than the hybridization enrgy. The crossover temperature 
then is much smaller than the Double-exchange energy favoring the existence of the bare spin-states.

This aspect pf the problem is absent in LiV and LiT because the two electronic states are both in the $t_{2g}$ manifold and the exchange energy is simply the 
hybridization energy.

{\em Acknowledgements}; Thanks are due to Anirvan Sengupta for discussions.

\end{document}